\pdfoutput=1
\documentclass{aastex631}

\usepackage{amssymb,amsmath}
\usepackage{soul}

\defcitealias{Brandt_2024a}{B24a}
\usepackage{gensymb}

\begin{document}

\title{Computing the Electronic Gain for Detectors Read Out Up-The-Ramp}

\author[0000-0003-2630-8073]{Timothy D.~Brandt}
\affiliation{Space Telescope Science Institute \\ 3700 San Martin Drive \\ Baltimore, MD 21218, USA}



\begin{abstract}

The electronic gain---the conversion between photoelectrons on a pixel and the digital number recorded to disk---gives physical units to an astronomical image and sets the relation between pixel value and photon noise.  This paper presents a new, likelihood-based approach to derive the gain from images taken up-the-ramp, where the detector is read out nondestructively many times before being reset.  Our method makes full use of the individual reads assuming an ideal detector subject to photon noise and Gaussian read noise.  We extend the method to account for slight nonlinearities in the relation between photoelectrons and measured counts.  We demonstrate that our likelihood-based approach provides a consistent (i.e.~asymptotically correct) and nearly unbiased estimator of the gain both with and without fitting for nonlinearity.  Finally, we apply this approach to a single detector from the Wide-Field Instrument on the Roman Space Telescope, and show how pixel-to-pixel gain variations describe much of the variations in pixel response seen in flatfield images.  Code to compute gain and regenerate figures in this paper is available at \url{https://github.com/RomanSpaceTelescope/SOCReferenceFileCode}.

\end{abstract}

\keywords{}


\section{Introduction} \label{sec:intro}

The electronic gain $g$, the number of electrons per DN (digital number) or ADU (analog-to-digital unit), is the conversion between a detector's raw reading as saved to disk and the number of photoelectrons actually received.  While flatfield images can correct the relative response of different pixels, they cannot distinguish variations in quantum efficiency (differing numbers of photoelectrons) from variations in gain (differing conversions to DN).  This is relevant when determining the uncertainty in the recorded value of a pixel, as shot noise produces an uncertainty of $\sqrt{n_e}$ in the number of photoelectrons $n_e$.  In order to derive the uncertainty on the raw counts in a pixel, we must know both the read noise and the gain.

The gain $g$ is typically measured using photon statistics via the {\it photon transfer curve} \citep{Janesick+Elliott+Collins+etal_1987,Janesick+Klaasen+Elliott_1987}.  If a detector is reset, illuminated, and read out many times, each time being illuminated identically, the scatter in the measured values will reflect both read noise and photon noise.  If it is then illuminated at some other level and read out many times, the read noise will typically be the same but the photon noise will differ.  By measuring the noise as a function of illumination, we can derive both the read noise and the gain for each pixel.

The photon transfer curve makes full statistical use of the data from a detector that is read out only once after each reset.  An example of such a detector is a CCD, in which the charge on a pixel is emptied as it is read out.  For a detector that can be read out nondestructively, however, the use of the photon transfer curve has important limitations.  Each segment of the ramp is subject to both read and shot noise, potentially providing much more information than is available from the total number of counts in the exposure.  A wide range of astronomical detectors are now read out up-the-ramp, with many nondestructive reads after each reset.  These include ground-based detectors like those on Subaru/CHARIS \citep{Groff+Chilcote+Kasdin+etal_2016,Brandt+Rizzo+Groff+etal_2017} and Gemini/GPI \citep{Macintosh+Graham+Ingraham+etal_2014}, the detectors on every instrument on JWST \citep{Doyon+Willott+Hutchings+etal_2023,Jakobsen+Ferruit+Oliveira+etal_2022,Rieke+Kelly+Misselt+etal_2023,Wright+Rieke+Glasse+etal_2023}, and the 18 detectors of the Wide-Field Instrument on the Roman Space Telescope \citep{Domber+Gygax+Aumiller+etal_2022,Schlieder+Barclay+Barnes+etal_2024}.  

This paper develops an alternative to the photon transfer curve for any detector read out up-the-ramp and subject to read noise, photon noise, and mild nonlinearity.  By using the full likelihood function of all of the reads, this method overcomes the limitations of the classical photon transfer curve and makes full use of the data.  The current approach retains the statistical assumptions of \cite{Janesick+Klaasen+Elliott_1987}: the noise is assumed to be solely due to photon and (Gaussian) read noise.  We make no attempt to disentangle different types of noise, e.g., as in \cite{Preece+Haefner_2022}.  

We organize the paper as follows.  Section \ref{sec:ptc_summary} summarizes the use of the photon transfer curve and its limitations with data read out up-the-ramp.  Section \ref{sec:likelihood} provides an overview of our likelihood-based approach.  Section \ref{sec:syntheticdata} validates the new approach on synthetic data, while Section \ref{sec:nonlinearity} introduces an extension of the method to account for first-order detector nonlinearities.  Section \ref{sec:implementation} discusses the implementation of our approach to gain measurement, and Section \ref{sec:roman_application} presents calculated gain maps for one of the detectors on the Wide-Field Instrument of the Roman Space Telescope.  We conclude with Section \ref{sec:conclusions}.

\section{Limitations of the Photon Transfer Function with Nondestructive Reads} \label{sec:ptc_summary}

The photon transfer curve \citep{Janesick+Klaasen+Elliott_1987} is based on the fact that read noise is independent of the number of photoelectrons, while shot noise is equal to the square root of the number of photoelectrons.  Assuming read noise and photon noise to be independent, the variance $\sigma^2$ in the measured accumulated charge on a pixel should be related to the read noise $\sigma_{\rm r}$, the accumulated charge $A$ in DN, and the gain $g$ by
\begin{equation}
    \sigma^2 = \sigma^2_{\rm r} + \frac{A}{g} .
    \label{eq:PTC}
\end{equation}
By illuminating a pixel at several different levels and measuring the variance of the recorded values at each illumination level, a line can be fit to $\sigma^2(A)$ in order to derive $g$ and $\sigma^2_r$.  For convenience in this section, we will rewrite Equation \eqref{eq:PTC} as
\begin{equation}
    S_i = R + \frac{A_i}{g}
    \label{eq:PTC_renamed}
\end{equation}
where $S_i$ is the variance of many exposures each at an illumination level corresponding to $A_i$ expected counts, and $R = \sigma_r^2$ is the read noise variance.

The use of Equation \eqref{eq:PTC_renamed} entails a number of assumptions, notably, the linearity of the detector and the uniformity of illumination through a series of exposures.  A more important limitation in the context of a detector read out up-the-ramp is that it refers to a single read difference between the beginning and the end of a ramp.  The differences between reads within a ramp can also be used, but the read noise in these differences is not independent.  For example, the differences between read 2 and read 1 and between read 3 and read 2 are correlated: read 2 is present in both read differences, with an identical realization of read noise.

To understand the importance of using the individual read differences, we will neglect read noise ($R \approx 0$) and derive the uncertainty on the values $S_i$ in Equation \eqref{eq:PTC_renamed}.  The uncertainty in the inverse gain $1/g$ will then scale with the uncertainty in $S_i/A_i$.  The maximum likelihood value of $S_i$ is
\begin{equation}
    S_i = \frac{1}{n} \sum (C_i - \langle C \rangle)^2
\end{equation}
where $C_i$ is the total number of counts in exposure $i$ and $\langle C \rangle$ is their expectation value.  Because this estimator for $S_i$ is the sum of the square of (nearly) Gaussian random variables, each of variance $\langle S_i \rangle$, the variance of this estimator of $S_i$ will have a variance equal to $2\langle S_i \rangle^2$ divided by the number of exposures $n$.  This is closely related to the variance of the $\chi^2$ distribution (itself the distribution of the sums of squares of Gaussian random variables).  Assuming $n_i$ measurements at each total number of expected counts $A_i$ and neglecting read noise, the uncertainty on the measured value of the variance $S_i$ will be 
\begin{equation}
    \sigma^2_{S,i}(\sigma_{\rm r}=0) \approx \frac{2 \langle S_i \rangle^2}{n_i} = 2 \frac{(A_i/g)^2}{n_i} .
\end{equation}
The quantity $S_i/A_i$, which will determine the measurement error in $1/g$, has an uncertainty 
\begin{equation}
    \frac{\sigma_{S,i}(\sigma_{\rm r}=0)}{A_i} \approx \frac{\sqrt{2}}{g\sqrt{n_i}} . \label{eq:error_scaling}
\end{equation}
Equation \eqref{eq:error_scaling} neglects uncertainty on $A_i$.  The relative uncertainty in $A_i$ scales as 
\begin{equation}
    \frac{\sigma_{A,i}}{A_i} \approx \frac{1}{\sqrt{n_i g A_i}} \label{eq:scaling_A}
\end{equation}
(with $gA_i$ giving units of photoelectrons), while the relative uncertainty in $S_i$ scales as 
\begin{equation}
    \frac{\sigma_{S,i}(\sigma_{\rm r} = 0)}{S_i} \approx \sqrt{\frac{2}{n_i}}. \label{eq:scaling_S}
\end{equation}
At low read noise and moderate and high illuminations $A_i$, $g A_i \gg 1$, and Equation \eqref{eq:scaling_A} is small compared to Equation \eqref{eq:scaling_S}. Equation \eqref{eq:error_scaling} then provides a good approximation to the uncertainty in $S_i/A_i$. 

Equation \eqref{eq:error_scaling} implies that as long as photon noise is large compared to read noise, the total number of counts on the detector does not matter, but only the number of measurements taken.  Because the uncertainty in $S_i$ is also proportional to the inverse gain, the relative precision on the gain $\sigma_g/g$ will scale as
\begin{equation}
    \frac{\sigma_g}{g} \approx \sqrt{\frac{2}{n_i}} .
    \label{eq:gain_precision}
\end{equation}
As a result, a higher number of photoelectrons does not help to determine the gain: the lever arm on the gain is larger, but the error bar on the variance increases to compensate.  The precision on the gain measurement comes purely from the number of measurements taken at a count rate large enough not to be read noise limited.  

In a long ramp, the differences between pairs of reads can function in the same role as independent exposures.  This greatly increases the constraining power of the data.  However, it comes at a cost: the differences between pairs of reads are correlated.  This correlation can be avoided by using only disjoint pairs of read differences: read 2 minus read 1, read 4 minus read 3, etc.  However, this discards half the data in order to avoid read noise covariance.  In order to use the full constraining power of a ramp, we must adopt a framework that accounts for this covariance.

\section{The Likelihood} \label{sec:likelihood}

The likelihood function naturally accounts for the covariance between reads, and includes the detector gain, read noise, and (with a small modification) nonlinearity.  We adopt the likelihood as the basis for the rest of our analysis.  The likelihood for a ramp, including covariance, assuming read noise and photon noise, is derived in \cite{Brandt_2024a} (hereafter \citetalias{Brandt_2024a}).  We refer to that paper for the full discussion of the approach to likelihood-based analyses of data taken up-the-ramp.  The count rate within a ramp is assumed to be constant and the detector is assumed to be ideal (subject only to read noise and photon noise).  The framework of \citetalias{Brandt_2024a} operates in the space of the differences of adjacent reads, e.g., read 2 minus read 1 and read 3 minus read 2, rather than in the space of reads directly.  This serves to make the covariance matrix tridiagonal and facilitates calculations.  

We begin with the likelihood function for a sequence of read differences from a detector read out up-the-ramp.  Equation (34) of \citetalias{Brandt_2024a} gives $\chi^2$, which may be identified with $-2 \log {\cal L}$, as
\begin{equation}
    -2 \log {\cal L} =
    \chi^2 = \left({\bf d} - a {\bf 1}\right)^T {\bf C}^{-1} \left({\bf d} - a {\bf 1}\right) \label{eq:chisq}
\end{equation}
where ${\bf C}$ is the covariance matrix of the data, ${\bf d}$ is a vector of read differences, $a$ is the count rate (in counts per read), ${\bf 1}$ is a vector of all ones, and $\log$ is the natural logarithm.  We assume here that all reads are saved individually, so we need not worry about the distinction between read and resultant (or group) made in \citetalias{Brandt_2024a} for when multiple reads are averaged together before being saved to disk.  The likelihood as given in Equation \eqref{eq:chisq} cannot quite be identified as the probability of the data given the model, as it lacks the normalization of the distribution given by the square root of the determinant of the covariance matrix.

We will now exponentiate Equation \eqref{eq:chisq} to write it as a likelihood and put the determinant of the covariance matrix back in so that it is normalized (up to factors of $\sqrt{2\pi}$).  The one thing that we will {\it not} account for here is that the covariance matrix is a function of the (unknown) count rate.  Adding in this dependence would make the analytic marginalization that follows impossible.  This limitation will be tested using Monte Carlo with synthetic data.  The likelihood (not log likelihood) becomes
\begin{align}
    {\cal L} = \frac{1}{\sqrt{{\rm det}\, {\bf C}}} \exp \left[ -\frac{1}{2} \left({\bf d} - a {\bf 1}\right)^T {\bf C}^{-1} \left({\bf d} - a {\bf 1}\right) \right] .
\end{align}

The likelihood is a function of the count rate, but it is also a function of the read noise and the gain.  This dependence arises through the covariance matrix
\begin{equation}
    {\bf C} = \frac{a}{g} {\bf C}_\gamma + \sigma^2_{\rm r} {\bf C}_{\rm r}
\end{equation}
(Equation (24) of \citetalias{Brandt_2024a}), where $\sigma_{\rm r}$ is in units of DN, $a$ is the count rate in units of DN/read, and and $g$ is the gain in units of $e^-/$DN.  

We will assume that it is sufficient to consider the covariance matrix to be a function of the read noise and the gain only, not the count rate.  In practice this means that the count rate will have to be reasonably well-determined by the ramp fit for ramps where photon noise is important.  This will occur in the limit of many reads per ramp, so the approach outlined here is only valid when applied to long ramps.  

We would like to have the likelihood as a function of the gain only, or jointly as a function of the gain and read noise.  Integrating the likelihood over the unknown count rate is much easier than integrating it over the read noise, so we will do that first.  We will assume that the integral over the count rate has limits of $\pm \infty$ rather than from zero to infinity.  This will make the integral much easier and only matters for count rates near zero where photon noise is insignificant; the covariance matrix ${\bf C}$ will always use a nonnegative count rate.  Denoting the read noise as $\sigma_{\rm r}$ and the gain as $g$, the integral to marginalize over the count rate becomes
\begin{align}
    {\cal L}\left(\sigma_{\rm r}, g \right) = \int_{-\infty}^\infty \frac{da}{\sqrt{{\rm det}\,{\bf C}}} \exp \left[ -\frac{1}{2} \left({\bf d} - a {\bf 1}\right)^T {\bf C}^{-1} \left({\bf d} - a {\bf 1}\right) \right] .
\end{align}
This is the ordinary Gaussian integral; its result is given by
\begin{equation}
{\cal L}\left(\sigma_{\rm r}, g \right) = \frac{1}{\sqrt{{\rm det}\,{\bf C}}} \sqrt{2\pi\sigma_a^2} \exp \left[ -\chi^2_{\rm best}/2 \right] 
\label{eq:likelihood_marginalized}
\end{equation}
where $\sigma_a$ is the uncertainty on the count rate and $\chi^2_{\rm best}$ is the minimum $\chi^2$; both of these are determined by the ramp fit described in \citetalias{Brandt_2024a}.  The determinant of ${\bf C}$ is also available: in the notation of \citetalias{Brandt_2024a}, it is the auxiliary quantity $\theta_n$ already computed in the course of deriving the inverse of the covariance matrix (Equation (40) of that paper).  As a result, ${\cal L}\left(\sigma_{\rm r}, g \right)$ may be computed straightforwardly using the existing ramp fitting software.  If we have many ramps, the total likelihood of $\sigma_{\rm r}$ and $g$ is the product of the likelihood computed for each ramp.  This likelihood may then be evaluated, mapped, and maximized as a function of assumed detector gain and read noise.

\section{Validation using Synthetic Data} \label{sec:syntheticdata}

To validate the approach developed in the previous section, we generate 100 ramps with a read noise of 10 electrons, a gain of 1 electron/DN, 30 reads/ramp, and count rates randomly varying from 0 to 100 electrons/read (each ramp having a different, random count rate).  We compute the likelihood using Equation \eqref{eq:likelihood_marginalized} as a function of $\sigma_{\rm r}$ and $g$: $\sigma_{\rm r}$ enters the covariance matrix straightforwardly according to the formalism of \citetalias{Brandt_2024a}, while the photon noise component of the covariance matrix must be divided by the gain.  The total likelihood is the product of the likelihood over all ramps, assuming that we allow each ramp to have a different count rate.  It will be more convenient to return to the log likelihood, so that we write
\begin{equation}
2 \log {\cal L}\left(\sigma_{\rm r}, g \right) = {\rm constant} + \sum_{\rm ramps} \left[ \log\left( {\rm det}\,{\bf C}\right) + 2\log \sigma_a -\chi^2_{\rm best} \right] .
\end{equation}
In the equation above, and throughout this paper, $\log$ refers to the natural logarithm.

\begin{figure*}
    \includegraphics[width=0.5\textwidth]{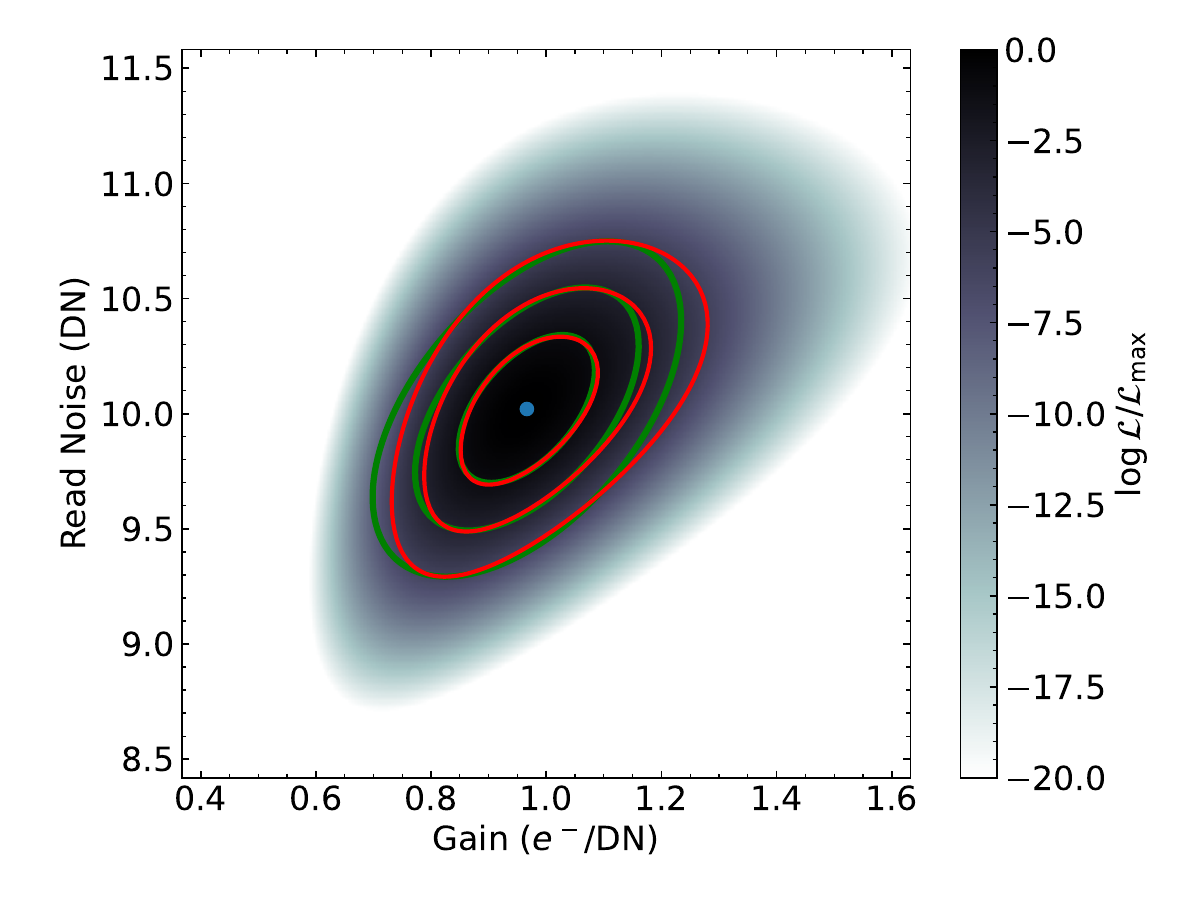}
    \includegraphics[width=0.5\textwidth]{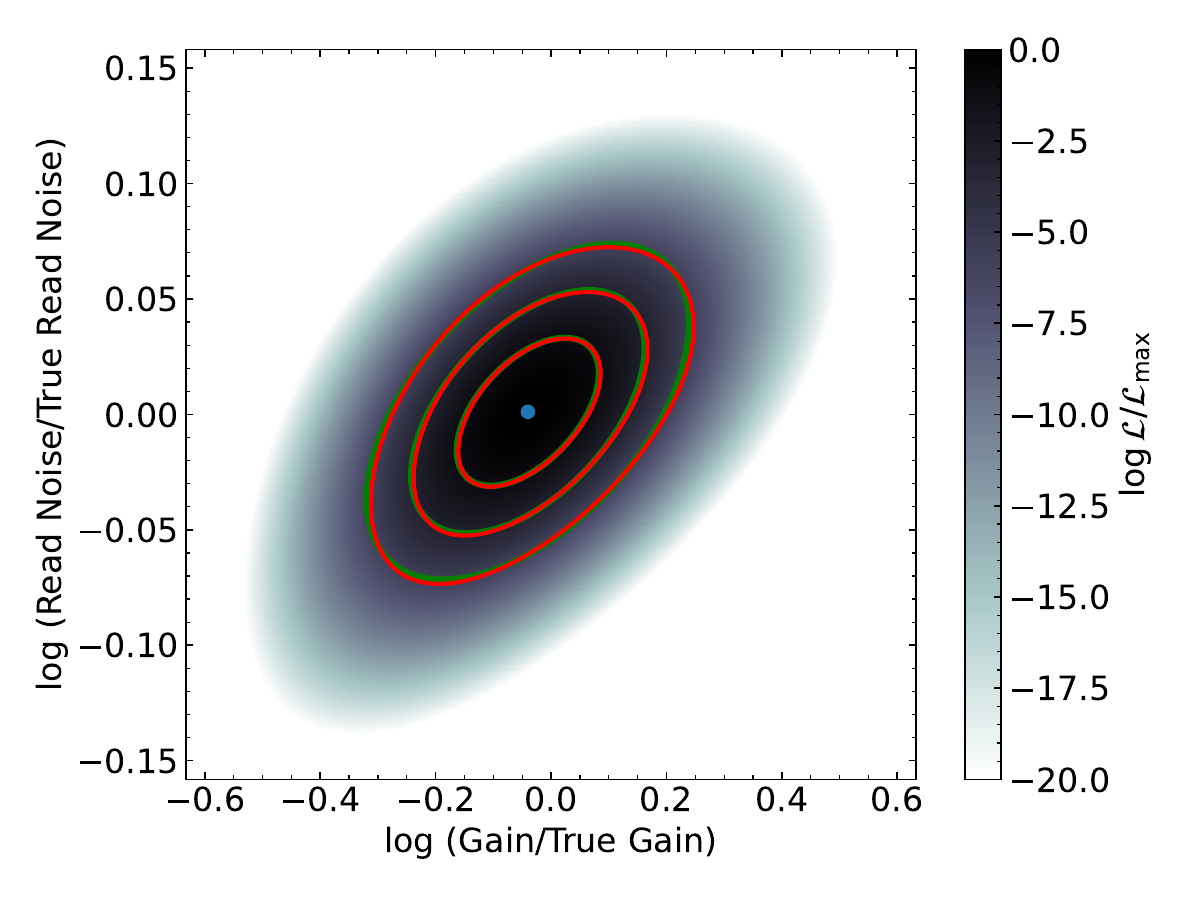}
    \caption{Log-likelihood as a function of gain and read noise, marginalized separately over 100 ramps and then combined over all of those ramps.  The red contours show $1\sigma$, $2\sigma$, and $3\sigma$ thresholds from $\Delta \chi^2$ thresholds.  The left panel interprets the likelihood as a probability density in gain and read noise; the right panel interprets the likelihood as a probability density in the logarithm of the gain and read noise.  The blue point and green contours are from fitting a quadratic form to the log likelihood at nine points near its peak.  When interpreting the likelihood as a probability density in log gain and log read noise, its logarithm is very accurately described by a quadratic form (equivalently, the likelihood is very nearly Gaussian).  In this case, shown in the right panel, the green contours---the $1\sigma$, $2\sigma$, and $3\sigma$ thresholds of a Gaussian likelihood described by the best-fit quadratic form---provide an excellent description of the likelihood.}
    \label{fig:likelihood_2d_test}
\end{figure*}

Figure \ref{fig:likelihood_2d_test} shows results for 100 synthetic 30-read ramps; the color is the log likelihood relative to its maximum value.  The red contours show the $1\sigma$, $2\sigma$, and $3\sigma$ levels computed using $\Delta \chi^2 = 2.3$, 6.2, and 11.8.  The contours in the left panel are not quite ellipses; the log likelihood is not quite a quadratic form near its maximum.  This means that the likelihood, if interpreted as 
\begin{equation}
{\cal L} \propto \frac{d^2 p}{d\sigma_{\rm r} dg} ,
\end{equation}
differs appreciably from a Gaussian.  

The form of the 2D probability density in the left panel of Figure \ref{fig:likelihood_2d_test}, combined with the fact that the read noise and gain should be positive, suggests interpreting the likelihood instead as
\begin{equation}
{\cal L} \propto \frac{d^2 p}{d\log\sigma_{\rm r} d\log g} . \label{eq:likelihood_interp2}
\end{equation}
This imposes the physical constraint that the gain and read noise must be positive.  The right panel of Figure \ref{fig:likelihood_2d_test} is the same as the left panel except that it interprets the likelihood according to Equation \eqref{eq:likelihood_interp2}.  The log likelihood now appears much closer to a quadratic form, i.e., the likelihood under this interpretation is much more accurately Gaussian.  The peak of the likelihood is slightly displaced from the true gain and read noise due to the realization of noise in the synthetic data.

To further test the Gaussianity of the likelihood, we fit a quadratic form to the log likelihood using nine points around its peak.  Figure \ref{fig:likelihood_2d_test} shows the $1\sigma$, $2\sigma$, and $3\sigma$ contours implied by this fit with green lines.  These are almost identical to the red contours computed using the log likelihood evaluated at thousands of points in order to derive contours at the standard $\Delta \chi^2$ offsets.  This shows that the likelihood, if interpreted as a probability density in $\log g$ and $\log \sigma_{\rm r}$, is accurately Gaussian.  We may fully determine it by evaluating the likelihood at just nine combinations of gain and read noise, and each of these evaluations requires fitting all ramps once.  We may then fit a quadratic form and use it to derive the best-fit gain and read noise together with their joint covariance matrix.  In practice, we take this one step further: we fit a quadratic form to determine the approximate location of the peak given an initial guess in gain and read noise, evaluate the likelihood at nine points around this approximate peak, and adopt the resulting best-fit quadratic form as our approximation of the true log likelihood.  The cost to fully determine the likelihood is then eighteen times the cost of fitting all of the ramps once.

\begin{figure}
    \centering
    \includegraphics[width=0.5\textwidth]{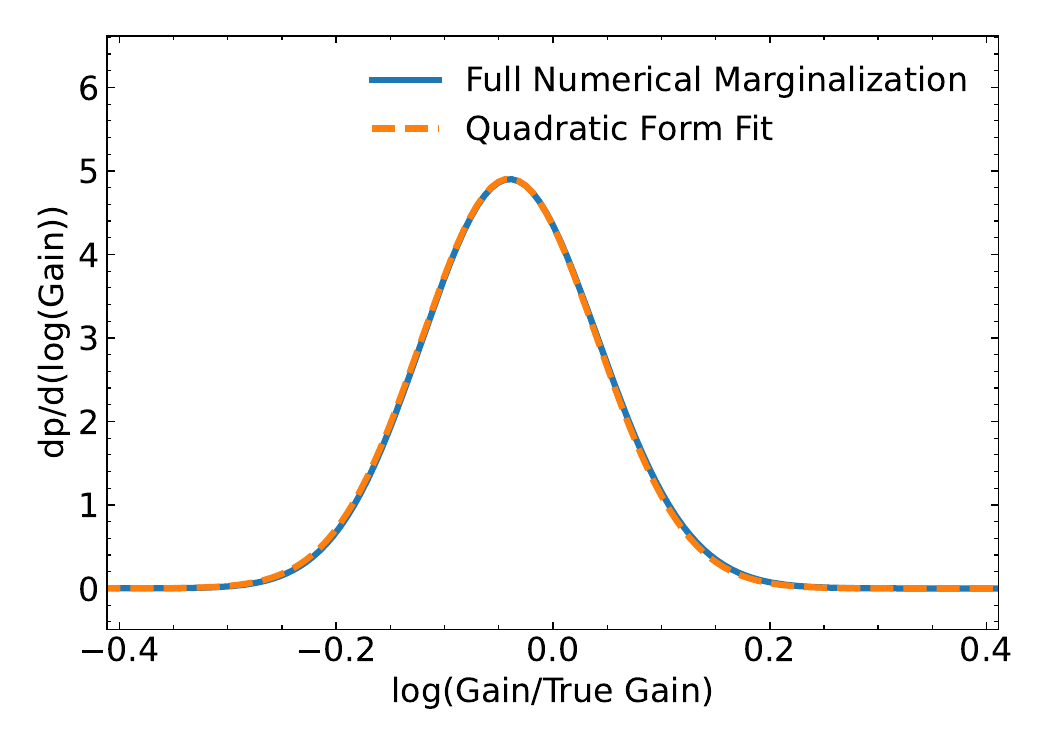}
    \caption{Probability distributions in $\log g$ for the same 100 ramps shown in the right panel of Figure \ref{fig:likelihood_2d_test} computed two ways: by a full numerical integral over the joint distribution shown in Figure \ref{fig:likelihood_2d_test} (solid blue line), and from the covariance matrix computed from the inverse of the quadratic form coefficients (orange dashed line).  The two probability distributions are almost indistinguishable.  The probability distribution of the gain for each pixel can thus be accurately computed using the approach outlined in this paper, at a cost of $\lesssim$20 times the cost of fitting the ramps used to determine the gain.}
    \label{fig:marginalized_pdists}
\end{figure}

While we might be interested in the joint probability distribution of the read noise and gain, we may also simply want the gain marginalized over the read noise.  In this case we need to integrate our two-dimensional likelihood over the read noise.  With the full covariance matrix of gain and read noise provided by the best-fit quadratic form, this is straightforward: the best-fit gain remains the same as in the joint distribution, while the uncertainty in the gain is the square root of the corresponding element of the covariance matrix.

We next evaluate the likelihood as a function of a single parameter--the gain--either by numerically integrating the likelihood or using the best-fit value and uncertainty provided by fitting a quadratic form.  The former approach is computationally prohibitive to use pixel-by-pixel, as it requires evaluating the likelihood thousands of times at each pixel.  Figure \ref{fig:marginalized_pdists} shows our results.  The solid blue line shows the probability distribution of $\log g$ computed by a brute-force marginalization of the full 2D distribution over $\log \sigma_{\rm r}$, while the red dashed line shows the Gaussian distribution derived from fitting a quadratic form to the log likelihood.  The two distributions are nearly identical: the marginalized posterior distribution of the gain is accurately Gaussian and may be characterized using a quadratic form fit to the log likelihood.  In our particular realization of noise, our recovered gain differs slightly from the true gain (though by less than one sigma).  The computational cost of our approach is linear in the number of pixels and the number of ramps.  Thanks to the scaling of the algorithms described in \citetalias{Brandt_2024a}, it is also linear in the number of reads.

\begin{figure*}
    \includegraphics[height=0.365\textwidth]{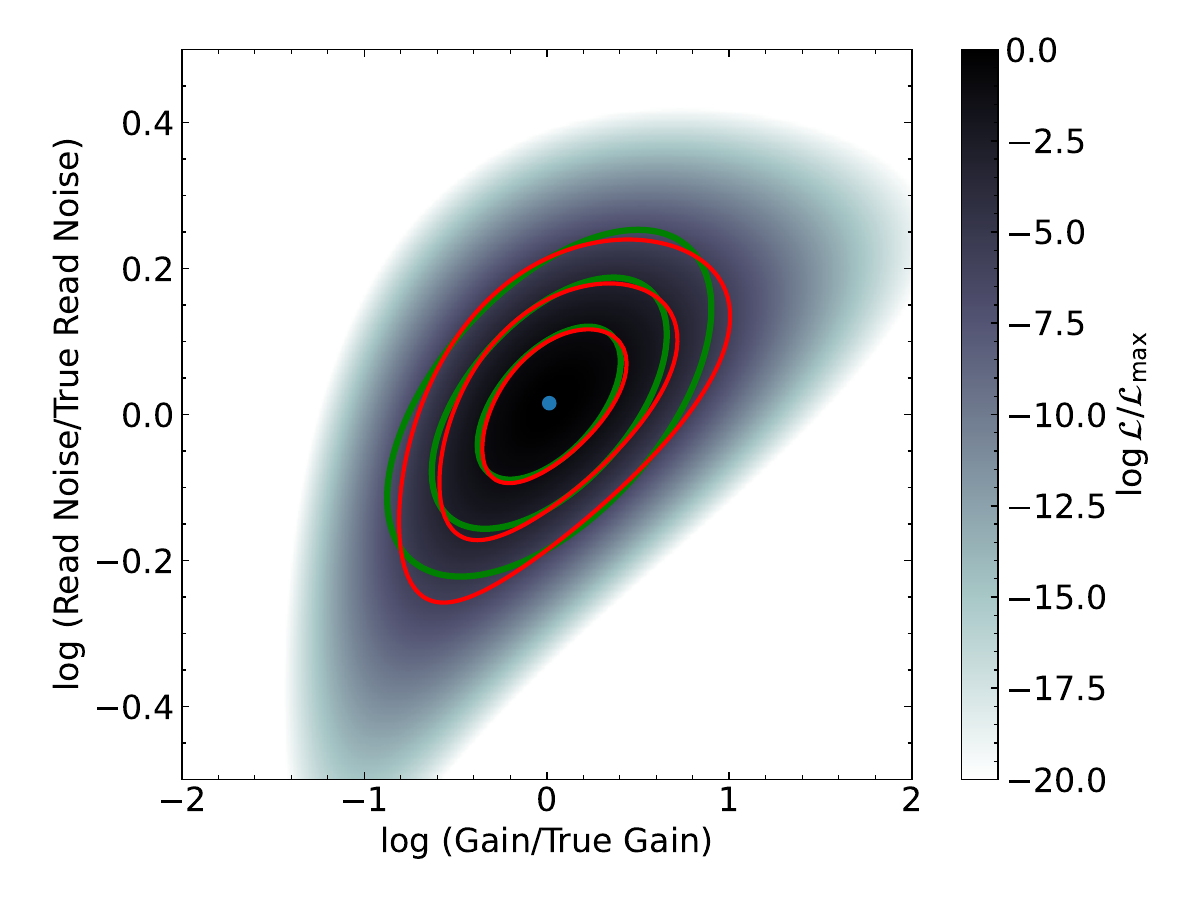}
    \includegraphics[height=0.365\textwidth]{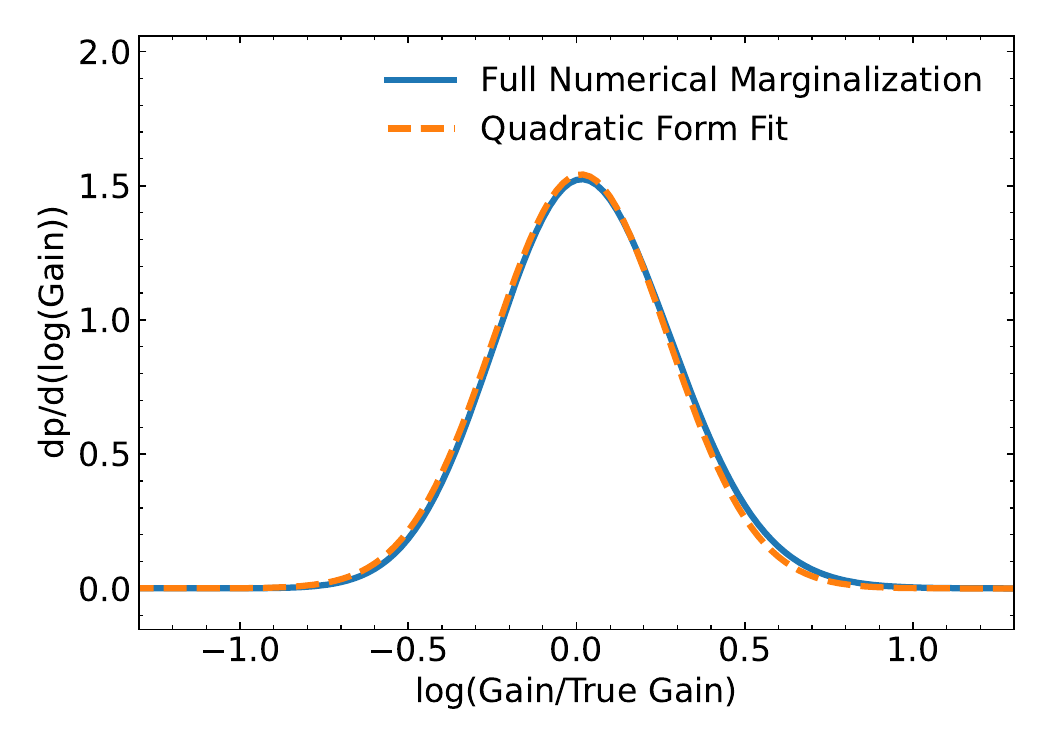}
    \caption{Left: same as the right panel of Figure \ref{fig:likelihood_2d_test}, but for just 10 ramps of 30 reads each.  The 10 ramps have random count rates ranging from 0 to 100 electrons/read.  The joint probability distribution remains accurately Gaussian, though not to the same degree as with 100 ramps.  Right: the marginalized probability distribution, computed as in Figure \ref{fig:marginalized_pdists}, likewise remains well-approximated by a Gaussian.}
    \label{fig:results_10ramps}
\end{figure*}

The previous analysis assumed 100 ramps each of 30 reads.  Finally, we repeat the analysis for just 10 ramps, to determine whether and how the likelihood begins to deviate significantly from Gaussianity.  Figure \ref{fig:results_10ramps} shows the results.  The uncertainties on read noise and gain are much larger with just 10 reads.  When interpreting the likelihood as a probability density in $\log g$ and $\log \sigma_{\rm r}$ (left panel), it remains nearly Gaussian, though not as accurately as with a larger number of reads or ramps.  The right panel of Figure \ref{fig:results_10ramps} shows that the one-dimensional likelihood marginalized over the read noise also remains accurately Gaussian even with just 10 ramps, for 300 total reads.  

It is possible that the read noise will be independently known.  Ideally it should be measured in the same configuration as the ramps used to fit for the gain. If the read noise is independently known, then fitting for the gain becomes a one-dimensional problem, and all of the discussion of fitting to a quadratic form will be replaced with fits to a parabola.  The computational cost will be correspondingly lower, though only by a factor of a few.  A parabola can be fully determined by three points while a two-dimensional quadratic form needs six.

We perform one additional check on the reliability of our results.  We generate 10,000 pixels' worth of ramps, with either 10 or 100 ramps per pixel, with each ramp having 30 reads and a random count rate from zero to 100 electrons/read.  We assume a true gain of 1 and then compute the $z$-score for each fit: the best-fit log(gain/truegain) divided by its standard error.  We then verify how closely the distribution of $z$-scores resembles a unit Gaussian.

\begin{figure*}
\includegraphics[width=0.5\textwidth]{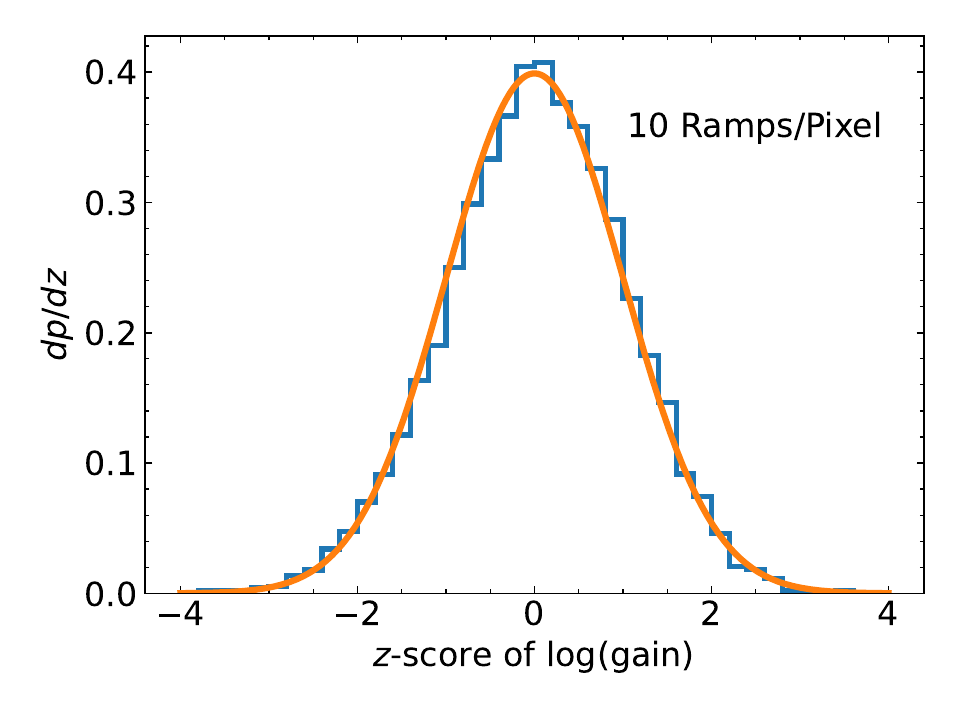}
\includegraphics[width=0.5\textwidth]{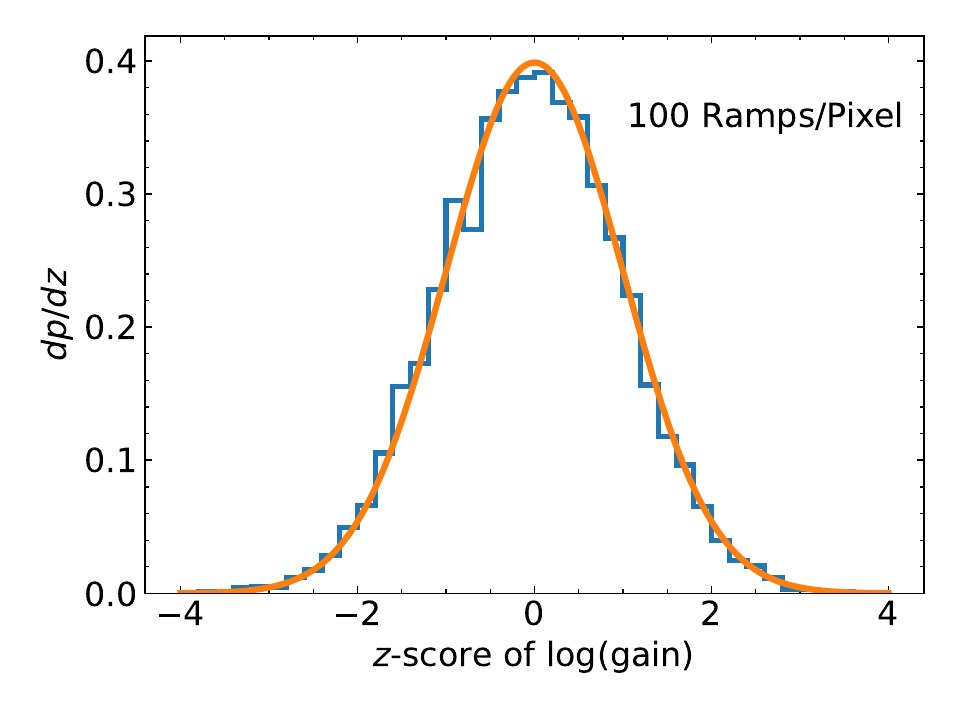}
    \caption{Distribution of $z$-scores (difference between true log(gain) and best-fit log(gain) divided by the standard error of the fit) compared to a unit Gaussian.  If the posterior probability distribution of the gain is reliable, the histogram and the unit Gaussian should match.  The two distributions are indeed an excellent match in both cases. This strongly suggests that if the noise model of Gaussian read noise plus Poisson photon noise is accurate, the method presented in this paper will produce reliable gain values and reliable uncertainties on those gain values. }
    \label{fig:zscores}
\end{figure*}

Figure \ref{fig:zscores} shows the results.  The distributions of $z$-scores of the best-fit gain values accurately match the theoretical unit Gaussians.  There is some deviation in the left panel, with just 10 ramps per pixel; this echos the small deviations from Gaussianity seen in the right panel of Figure \ref{fig:results_10ramps}.  With 100 ramps/pixel (for 3000 total reads per pixel), the distribution of $z$-scores is indistinguishable from a unit Gaussian.  This is evidence that our estimator is both unbiased (zero mean error) and consistent (zero asymptotic error) given sufficient data.  If we satisfy our assumptions of an ideal detector subject to independent, Gaussian read noise and Poisson noise on the total counts, then the approach outlined in this paper will provide accurate gain measurements making full use of the data.  

\section{Accounting for Nonlinearity} \label{sec:nonlinearity}

We will now expand the approach described in the previous sections to include a nonlinearity correction.  In the presence of strong nonlinearity, the gain will be expected to change substantially.  \cite{Rest+Hilbert+Leisenring+etal_2016} introduced a differential variant of the photon transfer curve to measure gain changes due to nonlinearity as a detector is read out up-the-ramp.  Here, we attempt to compute the gain at low count levels.  In this case, the nonlinearity correction is small and we can use a second-order Taylor expansion.  In practice, this means that the data used to determine the gain should only extend to a modest fraction of full well.  We model the nonlinearity by assuming that the measured voltage $V$ (converted to DN) is related to the incident photon rate $\gamma$ and integration time $t$ by
\begin{equation}
   \frac{1}{g} \gamma t = V_{\rm corr} = V + \alpha \left(V - V_0\right)^2
   \label{eq:nonlinearity_taylor}
\end{equation}
where $V_0$ is the reset voltage and $g$ is the gain.  This relationship is locally linear, but nowhere is it perfectly linear.  The gain, in this case, refers to the conversion between DN and photoelectrons for a small signal.  With the modification of Equation \eqref{eq:nonlinearity_taylor}, it is possible to fit the likelihood to $V_{\rm corr}$.  

In addition to fitting for the additional parameter $\alpha$, we may also need to account for read noise that follows a different nonlinearity correction from photoelectrons on a pixel.  This depends on the source of the nonlinearity.  If nonlinearity is due to the analog-to-digital converter (ADC), for example, then photon noise and read noise will both behave as expected in the corrected counts.  If nonlinearity is instead largely due to the relationship of charge and voltage on a nonideal capacitor, then read noise might behave as expected in the uncorrected counts, while photon noise would behave as Poisson noise in the corrected counts.  

For the present analysis we assume that photon noise scales in the same way as read noise, i.e., that photon noise is shot noise and read noise is white Gaussian noise in the corrected counts.  Correcting for nonlinearity then does not require modifying the covariance matrix for the data, but simply makes the covariance matrix a good description of the data.  Our assumption here is qualitatively supported for the Roman Space Telescope detectors by the fact that, as a pixel approaches saturation, the measured noise in DN (which includes read noise) becomes small.  We can then fit the likelihood using Equation \eqref{eq:nonlinearity_taylor} to transform the counts before fitting ramps.  Scaling the read noise by the correction factor applied to the counts makes little difference in the results of Section \ref{sec:roman_application}.

With the addition of a nonlinearity correction, the likelihood becomes a function of three parameters: gain, read noise, and the nonlinearity coefficient.  As before, we approximate the likelihood as a Gaussian in $\log g$ and $\log \sigma_{\rm r}$.  This yields three parameters, for 10 polynomial coefficients, requiring us to evaluate the likelihood at 2-3 times as many points as when we did not fit for the nonlinearity correction.  Avoiding bias is now more subtle, however: $\chi^2$ will change as a function of the nonlinearity coefficient, as an increase or decrease in the counts also changes the count differences.  With the values ${\bf d}$ now themselves dependent on a nonlinearity parameter $\alpha$, we must empirically test for biases in data with known true parameters.

There is an alternative to a three-dimensional likelihood evaluation with ${\bf d}$ depending on $\alpha$.  Intuitively, the best nonlinearity correction $\alpha$ should do the best job at making the counts increase linearity with time: it should maximize the consistency between the individual read differences within a ramp.  This will, in turn, maximize the derived gain.  We can optimize the likelihood as a function of gain and read noise at a series of nonlinearity corrections, and interpolate within this grid of nonlinearity coefficients $\alpha$ to find the $\alpha$ that produces the largest best-fit gain.  We then adopt this as our nonlinearity coefficient in order to derive the gain and read noise.  This approach will result in a slight positive bias on the gain, as it is usually possible to adjust nonlinearity parameters to fit a small amount of noise.  Assuming that the free choice of a nonlinearity parameter decreases $\chi^2$ by one, the gain will be biased by a fractional amount $\approx$$\sqrt{1/n_{\rm reads}}$.  We apply this small correction to the gains determined using the approach of maximizing gain as a function of nonlinearity.

We test both of the approaches above in the case of modest nonlinearity (a 10\% correction for the peak counts appearing in a data set).  For this purpose, we ensure that the true counts are exactly described by a quadratic function of the measured counts (which include both read noise and photon noise).  Our model of the noise and nonlinearity is thus perfect, and we are testing only the relative performance of two approaches to fitting the gain.  Both approaches are similar in their computational cost, being a few times as large as the optimization at fixed nonlinearity parameter.  The calculation assumes either 50 or 500 ramps, each of 50 reads with a read noise of 10\,e$^-$, comprising an even mixture of dark images with no electrons and photon-noise-dominated images with 100-200 electrons/read.  We compute 10,000 realizations of each set of ramps to show the distributions of best-fit log(gain), and normalize by our derived uncertainties to compute $z$-scores.

\begin{figure*}
    \includegraphics[width=0.5\textwidth]{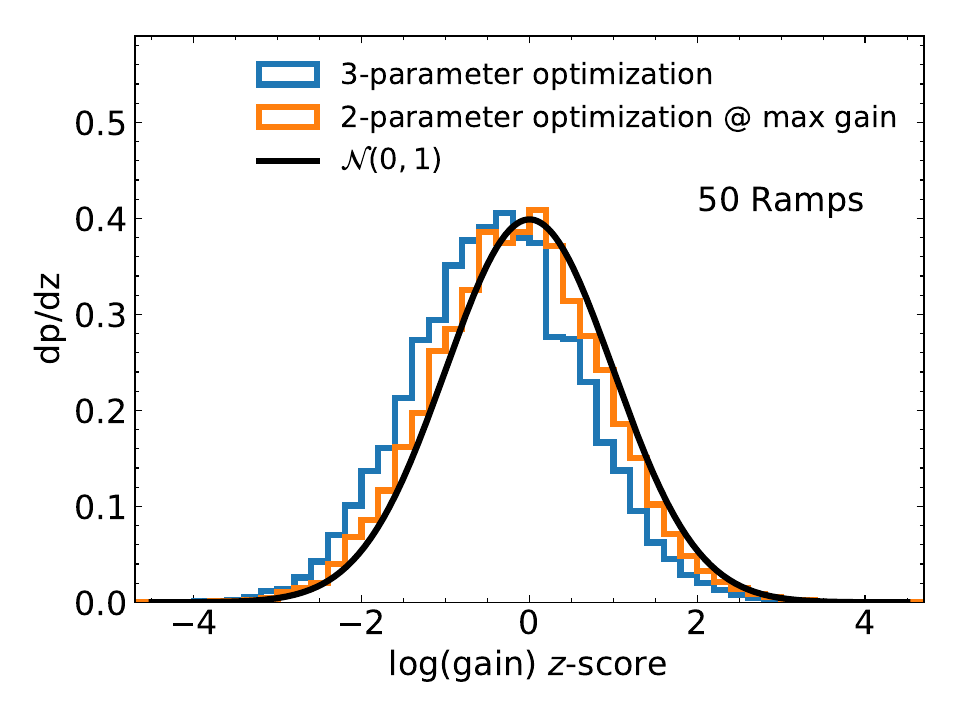}
    \includegraphics[width=0.5\textwidth]{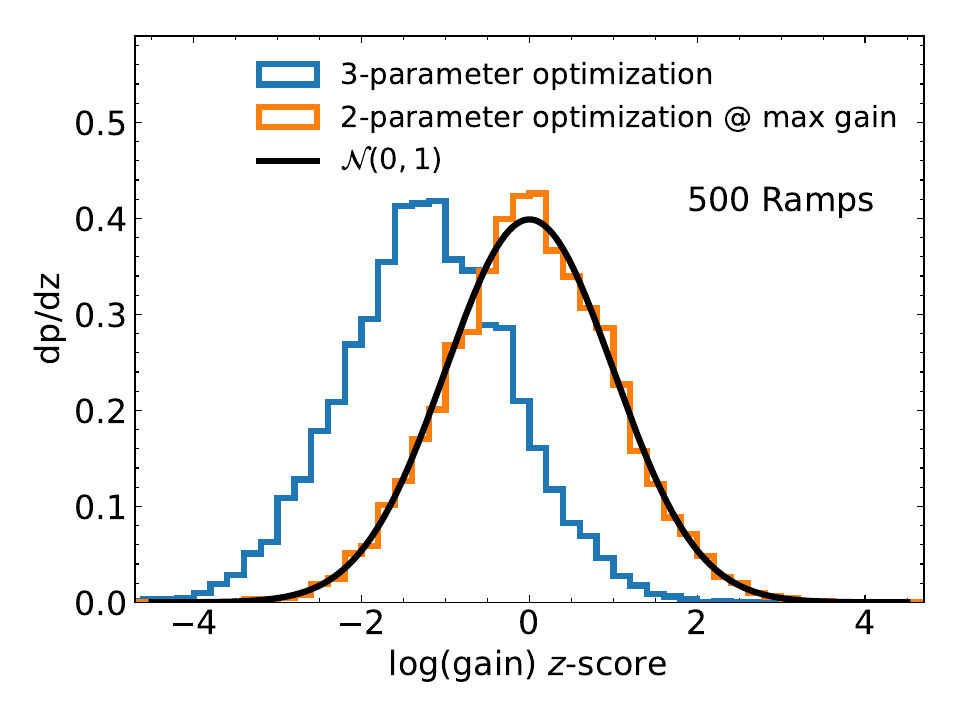}
    \caption{Distributions of $z$-scores of log(gain) using two approaches to fitting for nonlinearity.  The blue curves fit a three-dimensional likelihood in $\log g$, $\log \sigma_{\rm r}$, and nonlinearity coefficient $\alpha$, while the orange curves use a two-dimensional likelihood in $\log g$ and $\log \sigma_{\rm r}$ at the $\alpha$ that produces the largest gain.  In the latter case the derived gain is adjusted to account for the small bias expected from using the nonlinearity coefficient that maximizes the gain.  The left panel uses 50 ramps, each of 50 reads; the right panel uses ten times as much synthetic data. \label{fig:zscores_simultaneousfit}}
\end{figure*}

Figure \ref{fig:zscores_simultaneousfit} shows our results.  The approach of maximizing the derived gain appears to provide a consistent estimator.  After applying our bias correction, the best-fit gain approaches the true gain in the limit of infinite data.  With 500 total ramps the distribution of $z$-scores for many realizations of these ramps approaches a unit Gaussian.  This distribution is slightly displaced from zero with 50 ramps, indicating that our correction for bias is imperfect.  The approach of fitting a three-dimensional quadratic form produces a small, $\approx$3\% bias in the derived gain that becomes increasingly apparent as the error bars shrink with more data: this estimator is not consistent.  Motivated by Figure \ref{fig:zscores_simultaneousfit}, we adopt the gain-maximizing nonlinearity-fitting approach as our method of choice. 

Finally, we note that our analysis in this section assumed that our nonlinearity correction was perfect.  If there are nonlinearities that are not captured by a second-order Taylor expansion, the deviations from a constant count rate will be larger, and our inferred gain will be lower.  If the nonlinearity correction matches that used to reduce the data, however, our derived gains will still reflect the statistical consistency of a (corrected) ramp with a constant count rate.  They may still be useful for estimating the goodness-of-fit of a constant count rate, and for estimating an uncertainty on the best-fit count rate.

\section{Implementation} \label{sec:implementation}

Our implementation of the gain fitting method is largely built on the {\tt fitramp} module (\citetalias{Brandt_2024a}).  We have made a few minor modifications to suit the particular needs of this method.  These include returning the determinant of the covariance matrix, $\theta_n$ in the notation of \citetalias{Brandt_2024a}, for computing the normalization of the likelihood.  The second is a {\tt Cython} rewrite of the ramp fitting code itself.  {\tt fitramp} is vectorized so that many pixels may be handled simultaneously, making it efficient when fitting thousands of pixels at once.  If there are only a few hundred ramps for pixel, and each pixel is being treated one at a time, efficiency suffers.  Our use of {\tt Cython} gives a substantial performance improvement.

Some details of our implementation are motivated by its application to ground-based test data from Roman-WFI.  While selected results for Roman-WFI are presented in the following section, we summarize some of the relevant implementation details here.  {\tt fitramp} includes a full jump detection algorithm with an eye toward rejecting cosmic rays that are relatively common at L2, where JWST orbits.  Cosmic rays are far rarer in the ground-based data we use to determine gain.  We therefore apply a first pass on each pixel to search for jumps.  We first estimate the read noise as the median scatter of the differences in reads in the dark ramps.  We then mask read differences in the dark ramps that exceed a $5\sigma$ threshold. In the flatfield images, we adopt a fiducial gain of 1.85, appropriate to the Roman-WFI detectors, to estimate photon and read noise, and discard read differences that are more than $8\sigma$ discrepant from the median of a ramp.  For ground test data, this procedure masks a fraction $\sim$10$^{-5}$ of read differences.  Our fiducial gain could be tuned to values appropriate for other instruments.

Finally, we build an iterative optimization approach around our {\tt Cython} implementation of {\tt fitramp} to obtain the best-fit gain and read noise, along with their uncertainties.  We first use the dark ramps to set a prior on the read noise, fixing gain to 1 and nonlinearity to zero.  These choices are irrelevant, as dark ramps have nearly zero counts.  We fit a one-dimensional Gaussian to the likelihood of the dark ramps, and use its mean and variance to multiply the likelihood from the illuminated images by the likelihood from the darks.  We then turn to the illuminated data.  

We first choose three logarithmically spaced read noise values centered on the best-fit read noise from the dark ramps, and three logarithmically spaced gain values centered on a nominal value (typically near 2 for the Roman detectors).  We scale the spacing by the inverse of the square root of the number of samples, with logarithmic spacing of about 5\% in gain and 2\% in read noise for 200 ramps of 55 reads each.  This is slightly larger than the uncertainties that we ultimately derive.  We evaluate the likelihood on this grid for three different values of the nonlinearity coefficient centered on a typical value corresponding to a 1\% correction at about 10\% of full well.  

At each nonlinearity correction value, we compute the best-fit gain and fit a parabola to these three values to determine the best-fit nonlinearity parameter and gain.  We repeat this process in a finer grid about this initial best-fit nonlinearity parameter and gain.  Finally, we use this best-fit nonlinearity parameter to fit one more grid in read noise and gain.  As a final step, we decrease our final estimated gain slightly as described in Section \ref{sec:nonlinearity}.  We do not derive an uncertainty on the first-order nonlinearity coefficient.

The total cost to this approach is $\approx$60 times the cost of fitting all of the ramps once.  Each grid of gain and read noise has an evaluation cost nine times that of fitting a ramp once, and we perform this calculation at three different nonlinearity values two different times.  Finally, we compute the gain and read noise grid once more at the best-fit nonlinearity coefficient and near the best-fit gain.

\section{Application to Data from the Roman Space Telescope} \label{sec:roman_application}

In this section we present results from applying this analysis to data from one detector of the Wide-Field Instrument \citep[WFI,][]{Schlieder+Barclay+Barnes+etal_2024} of the Roman Space Telescope \citep{Perkins+Wollack+Content+etal_2024}.  We analyze Detector 9 in the Focal Plane Array, which lies to the lower-right of the center of the telescope's focal plane.  We derive the electronic gain using 200 dark ramps and 80 flatfield ramps.  All data were taken as part of instrument-level thermal vacuum (TVAC) testing campaigns\footnote{Data are available through in MAST in the Roman Integration and Test Archive (RITA) within the Calibration and Supplemental Search Interface (CASSI). For more information see \url{https://roman-docs.stsci.edu/data-handbook-home/accessing-wfi-data/the-roman-archive-in-mast/cassi}.}.  The dark ramps have a near-zero count rate, while for the flatfield ramps, the focal plane was illuminated as uniformly as possible using the Relative Calibration System (RCS, Rizzo et al.~in preparation).  All ramps that we use consist of 55 nondestructive reads.  The typical dark current that we observe is $\sim$0.01\,e$^-$\,pix$^{-1}$\,s$^{-1}$, so that the dark ramps overwhelmingly measure the read noise.  The flatfield images have typical count rates of $\sim$100\,DN\,read$^{-1}$, for a total accumulation of $\sim$5000\,DN in the exposure.  This is $\sim$10\% of full well, low enough that we expect a first-order nonlinearity correction to suffice.  

Hawaii-4RG detectors like those used in Roman-WFI have correlated noise with a characteristic $1/f$ power spectrum in time \citep{Rauscher+Fox+Ferruit+etal_2007,Rauscher_2015}.  Because of how the pixels on the detector are read out, a temporal correlation appears as a spatial correlation on the array.  Much of this noise may be removed by the use of reference pixels \citep{Moseley+Arendt+Fixsen+etal_2010}.  Before processing our data, we remove much of the $1/f$ noise using the Improved Roman Reference Correction \citep{2024romanrept..673B}.  After this step, the typical read noise is $\approx$5.3\,DN per read, or $\approx$7.5\,DN in a read difference.  With $\approx$100\,DN per read in our flatfield images and an expected gain near 2\,e$^-$/DN, that means that our images have similar contributions of read noise and photon noise to each read difference.  The core assumption of the approach in this paper is that read noise is uncorrelated from read to read.  The correlation of $1/f$ noise from pixel to pixel does not break this assumption.  So long as the $1/f$ noise retains a negligible uncorrected correlation from read to read (over a timescale of several seconds), our statistical assumptions are satisfied.  Spatially correlated noise could impart a small amount of spatial correlation to our recovered gain and read noise values.  We also perform a correction for highly structured, channel-dependent integral nonlinearity imparted by the analog-to-digital converter (ADC) using the approach of Brandt \& Perera (2025, submitted).  This is the nonlinearity visible in the lower panel of Figure 13 of \cite{Loose+Smith+Alkire+etal_2018}.  

\begin{figure*}
    \includegraphics[width=\textwidth]{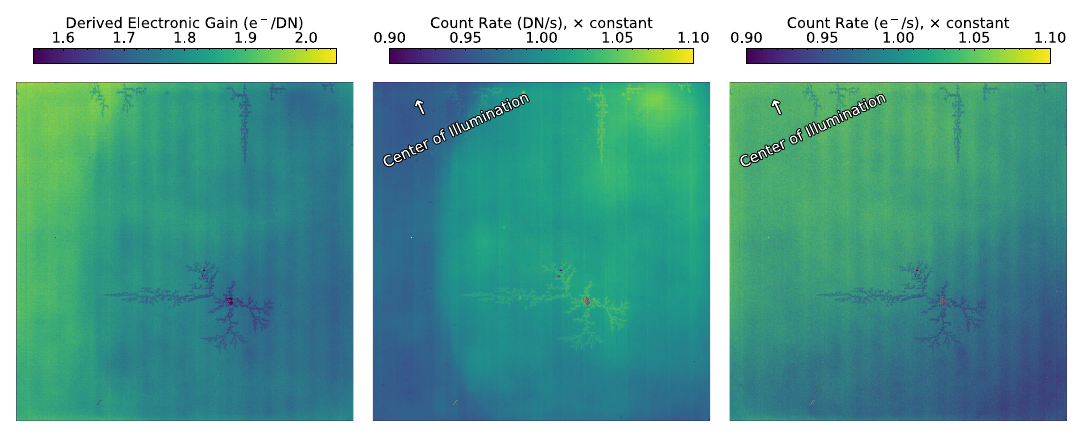}
    \caption{Left panel: derived electronic gain.  Middle panel: average count rate across flatfield images, corrected for nonlinearity, as measured in DN/s.  Right panel: the count rate in DN/s from the middle panel multiplied by the derived electronic gain to covert to e$^-$/s.  All count rates are divided by the median values in the images.  The center of illumination is above and to the left of the detector; we expect intensity to gradually drop as we move away from this direction. Gain variations explain much of the spatial structure in the flatfield count rates. \label{fig:full_flatfields}}
\end{figure*}

The left panel of Figure \ref{fig:full_flatfields} shows our derived electronic gain, in units of electrons/DN.  The median gain across the array is $\approx$1.8\,e$^-$/DN.  The gain shows significant structure and spatial variability, deviating by as much as $\approx$10\% from its median value.  The deviations are especially noticeable in feather-like features just below the center of the array and near the top of the array.  These regions, likely caused by deviations in the application of the epoxy coating, have significantly lower gains than the rest of the detector.  There is also a large region to the left of the detector where the gain is relatively high.  The 32 readout channels are visible in the image as faint vertical stripes.  The channels read out simultaneously; each one is 128 pixels wide.  Our median precision on the per-pixel gain is $\approx$0.065\,e$^-$/DN, or about 3.5\%.  With $\approx$4000 read differences in the 80 flatfield ramps, this is worse than the $\approx$2.2\% prediction of Equation \eqref{eq:gain_precision}.  The lower precision arises because the read noise is not negligible in our flatfield ramps, but is instead similar in magnitude to the photon noise.

The central panel of Figure \ref{fig:full_flatfields} shows the average best-fit count rate of 50 flatfield images taken in an optical configuration that enables maximum uniformity of illumination.  We have normalized the images by their median.  The direction of the center of illumination is noted: the intensity seen by the detector should increase somewhat in this direction.  Several features appear as the inverse of features seen in the gain map.  The region at the left of the detector now appears fainter, while the feather-like features appear brighter.  

The right panel of Figure \ref{fig:full_flatfields} shows the product of the left and middle panels: the illumination converted to units of electrons/s.  We again normalize this image by its median.  The inferred count rate in electrons/s lacks much of the spatial structure seen in the central panel.  The feathery features, which appeared brighter than the rest of the detector in DN/s, now appear fainter.  These pixels differ from the bulk of the detector in their electrical properties, recording higher values in DN despite being slightly less sensitive to photons.  The﻿ derived gain appears to ﻿change slightly between the first and last pixel read out in each (horizontal) row of a readout channel; this could reflect illumination-dependent settling as the readout progresses vertically from one row to the next. Adjacent channels are read out in alternating left-right order, so only 16 faint vertical bands are present from 32 readout channels. 

The right panel of \ref{fig:full_flatfields} shows an illumination that gradually decreases away from the center of illumination.  It has little spatial structure apart from feathery features with different electrical properties and small variations across readout channels.  The gain (left panel), which we infer solely from the statistical properties of the ramps, appears to explain much of the variation in the apparent illumination seen in the raw count rate images (middle panel).

Our approach neglects effects like burn-in and persistence.  These manifest as a count rate that depends on a pixel's recent illumination history.  They will cause a pixel's ramp to deviate from a straight line, and in turn, underestimate the gain.  We also perform no correction for interpixel capacitance (IPC), in which the voltage read out for one pixel is affected by the voltages of its neighbors.  IPC will cause an overestimate of the true gain, as the measured signal is smoothed slightly across neighboring pixels.  For our purposes, we are concerned primarily with the statistical properties of individual pixels and ramps rather than coupling between them.  Whether the user desires a gain that accounts for these effects or not depends on the intended usage.  The electronic gain derived in this paper is intended to provide an accurate statistical description of accumulated signal in a ramp.

\section{Conclusions} \label{sec:conclusions}

This paper has presented a new way of determining electronic gain for a detector read out nondestructively (up-the-ramp).  The approach is a generalization of the photon transfer method that uses photon statistics to infer gain.  The new method has several potential advantages over the classical photon transfer method:
\begin{enumerate}
    \item It uses all reads, fully utilizating the statistical power of the data.
    \item It adopts the correct covariance matrix for all of the reads.
    \item It does not assume that the true photon rate is the same between exposures.
\end{enumerate}
Our new method marginalizes over the (unknown) count rate in each pixel to derive a joint constraint on gain and read noise.  

We apply our new method both to synthetic data and to real data from TVAC testing of WFI on the Roman Space Telescope.  We find that our approach provides a consistent and very nearly unbiased estimator of the gain and that is also provides accurate uncertainties on the inferred gain.  Even in the presence of nonlinearity, we can recover the true gain if the nonlinearity is well-corrected by a second-order polynomial.  

The new gain calculation we present can be used to compute the gain separately for every pixel in a large format array, including all 18 Hawaii 4-RG detectors on Roman's WFI.  These gain values, in turn, may be used to weight the individual reads appropriately when fitting lines to the accumulated counts and in deriving uncertainties on the slopes of those lines.  Our gain values do not correct for effects like persistence, burnin, or interpixel capacitance.  Our goal here is to provide an accurate statistical description of ramps within individual pixels.

Finally, we demonstrate our likelihood-based gain measurement for one detector of the WFI instrument on Roman.  We derive the pixel-by-pixel gain to a median uncertainty of $\approx$3.5\%.  We then compare the flatfields in units of DN/read to the same flatfields in units of e$^-$/read.  We find that much of the spatial variation in the flatfield in units of DN can be accounted for by variations in gain, which are statistically detectable pixel-by-pixel.  These are especially evident in regions of the detector where the electrical properties are likely to be different to the deposition of the epoxy layer.  The gain method described and validated here will ultimately be used to derive the electronic gain maps for each of the eighteen detectors on WFI.  

{\it Software}: scipy \citep{2020SciPy-NMeth},
          numpy \citep{numpy1, numpy2},
          matplotlib \citep{matplotlib},
          Cython \citep{behnel2011cython},
          Jupyter (\url{https://jupyter.org/}).

\begin{acknowledgements}
  I thank an anonymous referee for comments that improved the manuscript.
\end{acknowledgements}

\end{document}